\documentclass[prl,twocolumn,citeautoscript]{revtex4}

\usepackage[T1]{fontenc}
\usepackage{txfonts}        
\usepackage{graphicx}
\usepackage{units}

\newcommand{\figsurface}      {./figure1.eps}
\newcommand{\figconductanceA} {./figure2a.eps}
\newcommand{\figconductanceB} {./figure2b.eps}
\newcommand{\figgating}       {./figure3.eps}

\begin{document}

\title{Ultrathin epitaxial graphite: 2D electron gas properties and a
  route toward graphene-based nanoelectronics.}

\author{Claire Berger}\altaffiliation[Permanent address: ]{CNRS-LEPES,
    BP166, 38042 Grenoble Cedex, France.}
\author{Zhimin Song}
\author{Tianbo Li}
\author{Xuebin Li}
\author{Asmerom Y. Ogbazghi}
\author{Rui Feng}
\author{Zhenting Dai}
\author{Alexei N. Marchenkov}
\author{Edward H. Conrad}
\author{Phillip N. First}
\author{Walt A. \surname{de Heer}}
\affiliation{School of Physics, Georgia Institute of Technology, Atlanta, GA 30332-0430}

\date{October 7, 2004}

\begin{abstract}
  We have produced ultrathin epitaxial graphite films which show
  remarkable 2D electron gas (2DEG) behavior. The films, composed of
  typically 3 graphene sheets, were grown by thermal decomposition on
  the (0001) surface of 6H-SiC, and characterized by surface-science
  techniques. The low-temperature conductance spans a range of
  localization regimes according to the structural state (square
  resistance \unit[1.5]{k$\Omega$} to \unit[225]{k$\Omega$} at
  \unit[4]{K}, with positive magnetoconductance). Low resistance
  samples show characteristics of weak-localization in two dimensions,
  from which we estimate elastic and inelastic mean free paths. At low
  field, the Hall resistance is linear up to \unit[4.5]{T}, which is
  well-explained by $n$-type carriers of density
  \unit[$10^{12}$]{cm$^{-2}$} per graphene sheet.  The most
  highly-ordered sample exhibits Shubnikov - de Haas oscillations
  which correspond to nonlinearities observed in the Hall resistance,
  indicating a potential new quantum Hall system. We show that the
  high-mobility films can be patterned via conventional lithographic
  techniques, and we demonstrate modulation of the film conductance
  using a top-gate electrode. These key elements suggest electronic
  device applications based on nano-patterned epitaxial graphene
  (NPEG), with the potential for large-scale integration.
\end{abstract}


\maketitle



The exceptional electronic transport properties of low-dimensional
graphitic structures have been amply demonstrated in carbon nanotubes
and nanotube-based transistors. Ballistic transport has been observed
up to room temperature \cite{Frank98,Poncharal02,Liang01}, and quantum
interference effects at cryogenic temperatures
\cite{Tans97--0,Bachtold99,Schonenberger99}. Simple nanotube transistors
\cite{Tans98--0,Martel98}, and interconnected logic gates
\cite{Bachtold01} have been demonstrated, which rely on the ability to
control the nanotube conductance via an electrostatic gate. The basic
transport parameters of these devices are so compelling that nanotubes
are considered to be a candidate material system to eventually
supplant silicon in many electronic devices.

An under-appreciated fact is that most electronic properties of carbon
nanotubes are shared by other low-dimensional graphitic structures.
For example, planar nanoscopic graphene ribbons (i.e. ribbons of a
single sheet of graphite) have been studied theoretically
\cite{Wakabayashi01--0,Nakada96--1}, and they exhibit properties that are
similar to nanotubes. Graphene ribbons with either metallic or
semiconducting electronic structure are possible, depending on the
crystallographic direction of the ribbon axis \cite{Wakabayashi01--0}.
Thus, if suitable methods were developed to support and align graphene
sheets, it would be possible to combine the advantages of
nanotube-like electronic properties with high-resolution planar
lithography to achieve large-scale integration of ballistic devices.
An essential difference between nanotubes and planar graphene ribbons
is the presence of dangling bonds at the edges. Normally these would
be hydrogen-terminated, with little influence on the valence
electronic properties. However, edge atoms could be passivated with
donor or acceptor molecules, thus tuning the electronic properties
without affecting the graphitic backbone of the device.

This Letter presents recent results \cite{Berger04} that show the
two-dimensional nature of electrical transport in ultrathin graphite
(multilayered graphene) grown epitaxially on SiC(0001). 6H-SiC is a
large bandgap (\unit[3]{eV}) semiconductor, which provides an
insulating substrate at temperatures below \unit[50]{K} for the
$n$-type (nitrogen) doping employed here. We use magnetoconductance
measurements and the physics of weak-localization to determine
transport parameters of the graphite 2D electron gas (2DEG), and we
show that the character of the magnetotransport/localization spans a
wide range of behaviors, depending on the amount of disorder in the
film or substrate. Quantum oscillations in the magnetoconductance and
in the Hall resistance are found for the most ordered sample. The
character of these features suggests that the quantum Hall effect
could be observed at lower temperatures, higher fields, or in
ultrathin graphite films of only slightly higher mobility.  To our
knowledge, these are the first transport measurements on oriented and
patterned graphite films of only a few monolayers thickness (hence
``graphene'' films), although related transport experiments have been
done on thicker (65--100 graphene layers) free-standing graphite
microdisks, which were nano-patterned by focused-ion-beam (FIB)
lithography \cite{Dujardin01}.

Given the large mean free paths measured in high-quality graphites
\cite{Kaburagi96}, the unusual electronic dispersion of graphene, and
the fact that the carriers lie near an air-exposed surface, this
unique 2DEG system holds great scientific potential. Furthermore, with
sufficiently high-quality material, ballistic and coherent devices
analogous to nanotube designs \cite{Javey03} would be possible.  This goal
requires that the epitaxial graphene can survive the processing
necessary for creation of submicron ribbons
\cite{Wakabayashi01--0,Nakada96--1}, and that the 2DEG can be gated
electrostatically. Below we also demonstrate these critical elements
for the realization of electronic devices based on nano-patterned
epitaxial graphene (NPEG).


Ultrathin epitaxial graphite films were produced on the Si-terminated
(0001) face of single-crystal 6H-SiC by thermal desorption of Si
\cite{Bommel75,Charrier02,Forbeaux98--0,SampleNote}.  After surface
preparation by oxidation \cite{Cho03} or H$_{2}$ etching
\cite{Ramachandran98}, samples were heated by electron bombardment in
ultrahigh vacuum (UHV; base pressure \unit[$1\times 10^{-10}$]{Torr})
to \unit[$\sim$1000]{$^{\circ}$C} in order to remove the oxide (some
samples were oxidized/de-oxidized several times to improve the surface
quality). Scanning force microscopy images showed that the best
initial surface quality was obtained with H$_2$ etching (sample A).
After verifying by Auger electron spectroscopy (AES) that the oxide
was removed, samples were heated to temperatures ranging from
\unit[1250]{$^{\circ}$C} to \unit[1450]{$^{\circ}$C} for 1--20 min.
Under these conditions, thin graphite layers are formed
\cite{Bommel75,Charrier02,Forbeaux98--0}, with the layer thickness
determined predominantly by the temperature.  Multilayered-graphene
film thicknesses were estimated by modeling the ratio of measured
intensities in the Si 92-\unit{eV} and C 271-\unit{eV} Auger peaks
(\unit[3]{keV} incident energy) \cite{Li04,Tanuma91--0,Tanuma91--1}.

\begin{figure}
  \includegraphics[width=\columnwidth]{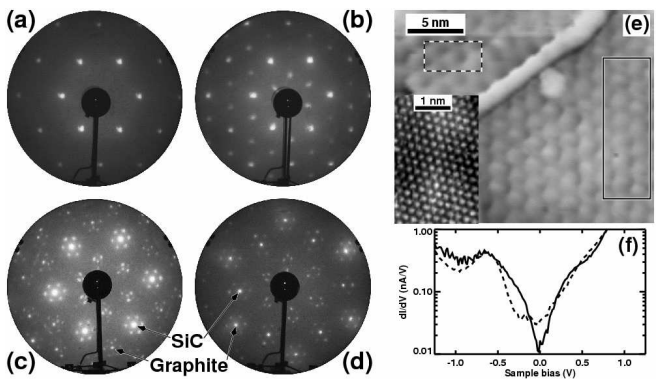}
  \caption{\label{fig:surface} (a)-(d) LEED patterns from
    Graphite/SiC(0001). The sample was heated several times to
    successively higher temperatures. (a) \unit[1050]{$^{\circ}$C} for
    10 min. Immediately after oxide removal, showing SiC $1 \times 1$
    pattern at \unit[177]{eV}. AES C:Si ratio 1:2. (b)
    \unit[1100]{$^{\circ}$C}, 3 min.  The $\sqrt{3}\times\sqrt{3}$
    reconstruction is seen at \unit[171]{eV}.  AES ratio 1:1.9.  (c)
    \unit[1250]{$^{\circ}$C}, 20 min.  \unit[109]{eV} pattern showing
    diffracted beams from the $6\sqrt{3} \times 6\sqrt{3}$ unit cell.
    Examples of first-order SiC and graphite spots are marked.  Note
    the surrounding hexagons of ``$6 \times 6$'' spots.  AES C:Si
    ratio 2:1 (\unit[$\sim 1$]{ML} graphite). (d)
    \unit[1400]{$^{\circ}$C}, 8 min.  \unit[98]{eV} LEED pattern.  AES
    ratio 7.5:1 (\unit[$\sim 2.5$]{ML} graphite).(e) STM image of a
    surface region of the sample described in Fig.~\ref{fig:surface}d.
    Inset: Atomically-resolved region (different sample, similar
    preparation).  (f): $dI/dV$ spectra (log scale) acquired from the
    regions marked with corresponding line types in the image at top.
    The solid line is an average of 396 spectra at different
    positions, the dashed line an average of 105. With a few
    ``glitchy'' exceptions, individual spectra in each region showed
    negligible variation from the average $dI/dV$ shown. }
\end{figure}
Figures \ref{fig:surface} (a)--(d) show low energy electron
diffraction patterns (LEED) at different stages during the growth of a
2.5-monolayer (ML) graphite film grown in-situ.
Figure~\ref{fig:surface}e displays an STM image from the sample
obtained after stage (d). The image reveals a distinct $6 \times 6$
corrugation of the overlayer \cite{Tsai92} and a raised region along a
step on the surface.  This modulation has been previously attributed
to variations of the interlayer interaction arising from Moiré
coincidences between the graphite and SiC lattices within a
fundamental $6\sqrt{3} \times 6\sqrt{3}$ surface unit cell
\cite{Bommel75,Tsai92,Starke98--1,Owman96--0}.  LEED confirms that the
graphene sheets register epitaxially with the underlying SiC, as shown
in Figs.~\ref{fig:surface}c and \ref{fig:surface}d. The mean height
difference between terraces (\unit[0.25]{nm}), indicates that the step
in Fig.~\ref{fig:surface}e is a bilayer step in the SiC substrate.
Terrace sizes (corresponding to a single $6 \times 6$ domain) are
found by STM to be up to several hundred nanometers in extent.
Preliminary high-resolution LEED studies indicate that the graphene
layers are strained in-plane by 0.3--0.5\%, with a mean structural
coherence length of greater than \unit[20]{nm}.

Also shown in Fig.~\ref{fig:surface}f are derivative tunneling spectra ($dI/dV$
vs.\ $V$) acquired within the respective boxed regions of the
image.  The $dI/dV$ spectrum obtained from the lower terrace
(solid line) is consistent with that of a zero-gap semiconductor, as
found typically for bulk graphite. On the upper terrace, the $6 \times 6$
domain images somewhat differently, and the $dI/dV$
spectrum (dashed line) displays a region of constant, finite
conductance around the Fermi energy (zero bias).  Spectral shapes are
very uniform within each $6 \times 6$ domain.  The $dI/dV$
curves show that the electronic properties of the film are not
entirely homogeneous. This may relate to differing lateral registry
(i.e. not orientational) of the graphite on the SiC substrate, or
electron confinement within $6 \times 6$ domains.  $dI/dV$
spectra acquired over the buckled region at the step edge are nearly
identical to those found on the upper terrace, suggesting that the
graphite layer remains continuous over the step.

\begin{figure}
\resizebox{\columnwidth}{!}{%
  \includegraphics[scale=1.0]{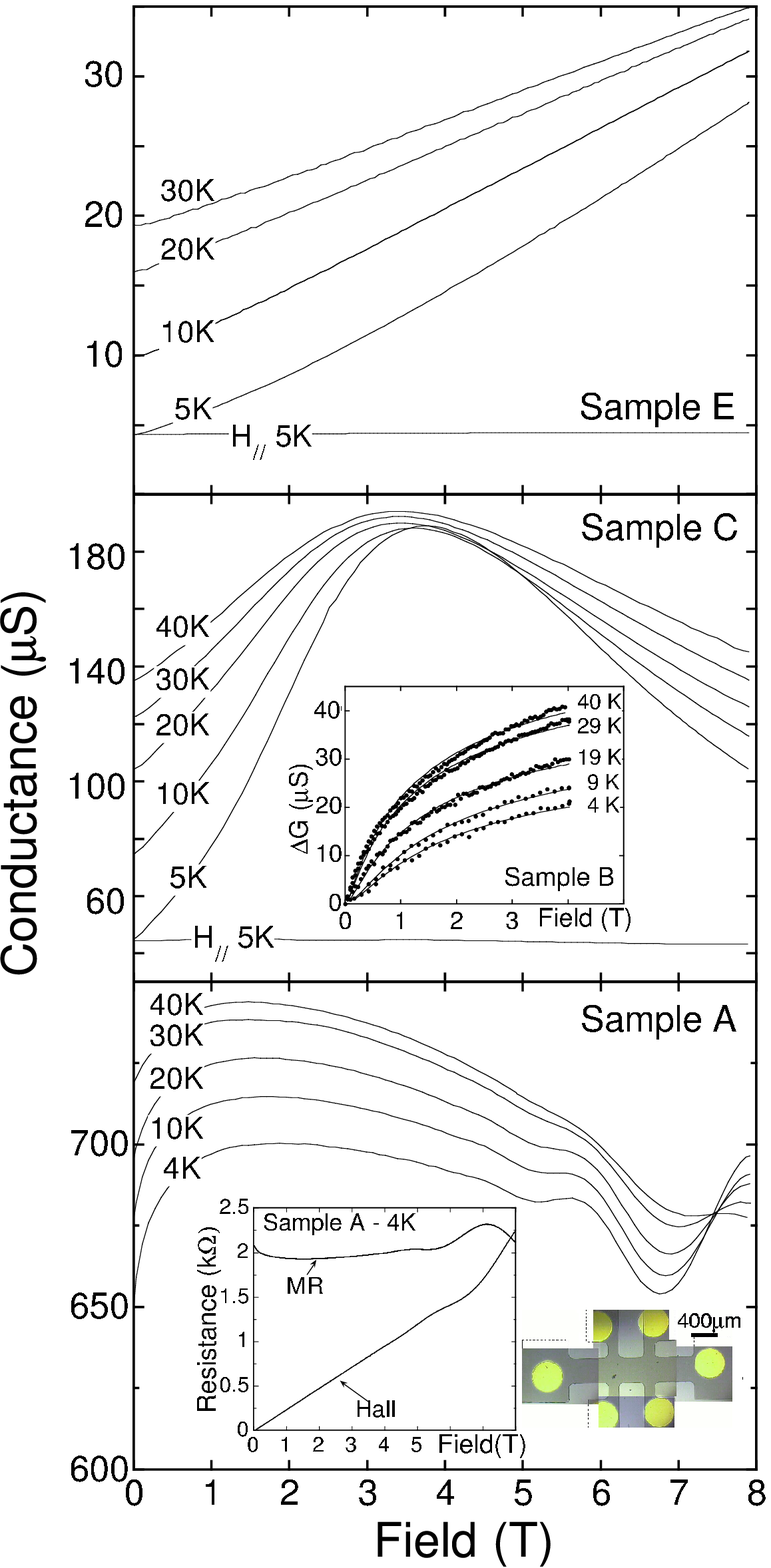}~~~
  \raisebox{1.08\height}{\includegraphics[scale=0.9]{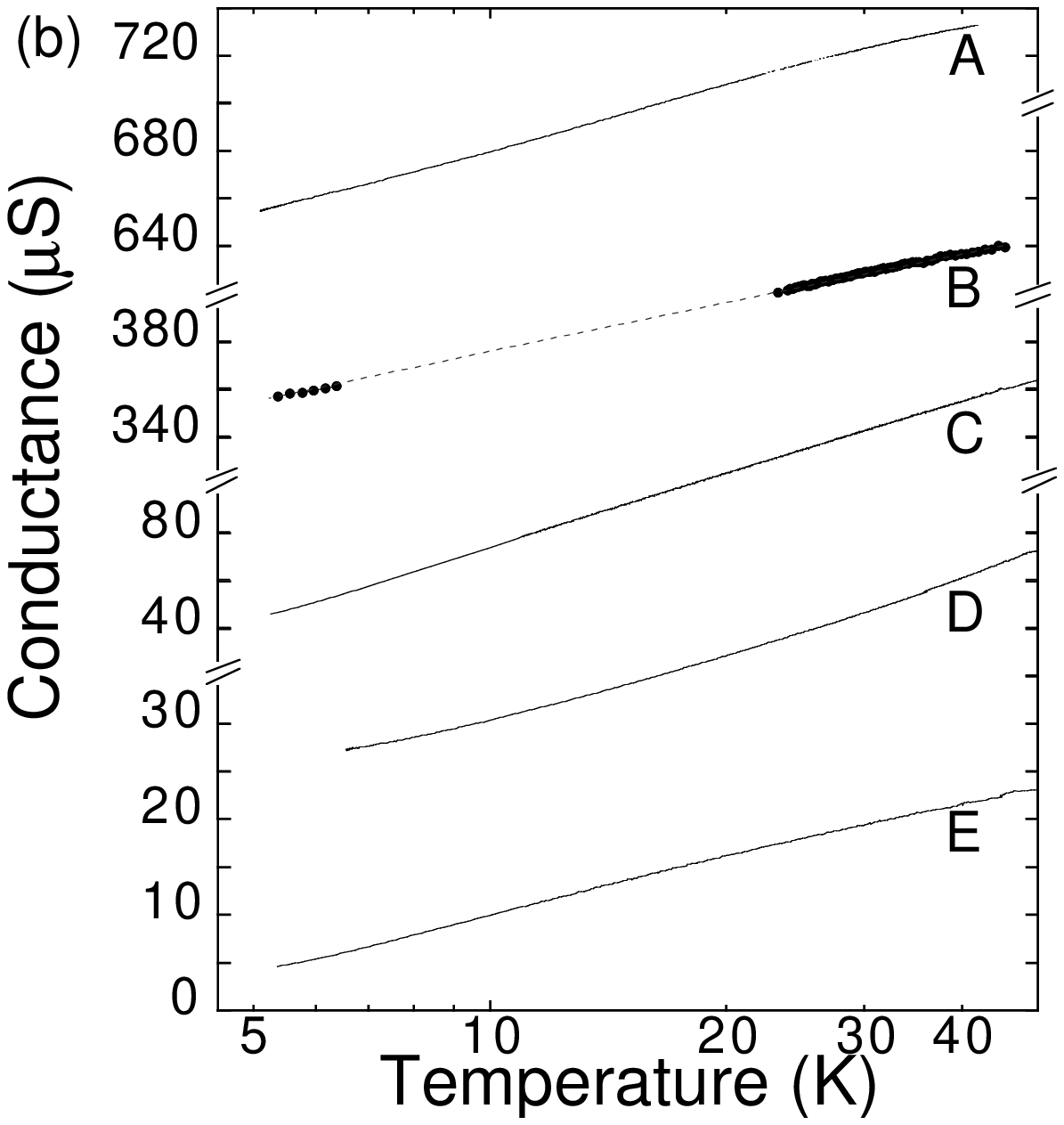}} }
\caption{\label{fig:conductance} (a) Magnetoconductance $G(H)$ for
  samples E, C and A in perpendicular field for temperatures as
  indicated. Also shown for samples E and C, the in-plane
  magnetoconductance at \unit[5]{K}.  Note the strong anisotropy
  between the perpendicular and in-plane field configuration. In
  sample C, the conductance goes through a maximum at about the same
  value of field and conductance for all $T$.  In sample A
  Shubnikov-de Haas oscillations develop at high magnetic field.
  Center-panel inset : $G(H)-G(H=0)$ for sample B.  Solid curve are
  fits to the weak localization theory with no spin-orbit, yielding
  elastic scattering length $\ell_e\approx \unit[15]{nm}$ and
  inelastic scattering length $\ell_i(4K)\approx \unit[100]{nm}$.
  Lower-panel inset : Hall resistance $R_{xy}=V_{xy}/i$ and
  magnetoresistance $R_{xx}$ as a function of magnetic field at 4K for
  sample A. Below \unit[4.5]{T}, $R_{xy}$ is remarkably linear,
  indicative of $n$-type carriers. At high magnetic field, the
  oscillations coincide with the oscillations in the
  magnetoresistance. (b) Square conductance as a function of
  temperature for samples of various conductance (samples A, B, C, D
  and E from top to bottom). Note that the conductance scale is
  $\times 4$ for samples D and E, compared to the scale for sample
  A-C. Samples A, B and C show a $\ln T$ form in this decade of
  temperature, characteristic of 2D weak localization.  }
\end{figure}

DC and low-frequency AC conductance measurements were made for
temperatures $T=\textrm{0.3--50}]{K}$, and for magnetic fields $H$
from \unit[0--8]{T} on graphite films with thicknesses of typically 3
graphene sheets (see Table~\ref{tab:samples}). For Hall-effect
measurements, samples were defined using standard optical lithography
(photoresist coating, plasma etching, photoresist removal via
solvents). Four contacts were painted with silver paste directly on
the surface or on evaporated Pd-Au pads on mm-size samples. For
samples B, C and E, the voltage probe distance $d_V$ is \unit[2]{mm}.
For the Hall bar samples A and D $d_V = 600$ and \unit[300]{$\mu$m}
respectively (see photo inset, Fig.~\ref{fig:conductance}a).  Reported
values below are the square conductance $G$.
\begin{table}
  \begin{center}
    \begin{tabular}{crr@{~}lr@{~}lr@{~}l}
      Sample &  \multicolumn{1}{l}{C:Si} & \multicolumn{2}{l}{Thickness}&\multicolumn{2}{l}{$R_{4K}$} & \multicolumn{2}{l}{Mobility}\\ \hline\hline
      A & \hfill 10~       & \hfill 3    &ML& \hfill 1.5 &k$\Omega$& \hfill 1100 &cm$^{2}$/Vs\\
      B & \hfill $\infty$~ & \hfill $>5$ &  & \hfill 2.2 & & \hfill      & \\
      C & \hfill  9~       & \hfill 3    &  & \hfill 22  & & \hfill      & \\
      D & \hfill 10~       & \hfill 3    &  & \hfill 33  & & \hfill 15   & \\
      E & \hfill  9~       & \hfill 3    &  & \hfill 225 & & \hfill      & \\
      F & \hfill  7~       & \hfill 2.5  &  & \hfill     & & \hfill      & \\
    \end{tabular}
  \end{center}
  \caption{\label{tab:samples} Sample properties. Ratio of intensities
    in the C(\unit[271]{eV}) and Si(\unit[92]{eV}) AES peaks,
    calculated thickness in graphene monolayers, square resistance at
    \unit[4]{K}, and mobility (where measured). }
\end{table}

The 2D nature of electrical transport in the film is vividly
demonstrated in Fig.~\ref{fig:conductance}a by the large anisotropy in
the magnetoconductance [$\textrm{MC}=G(H)$]: For $\mathbf{H}$
perpendicular to the graphene plane the differential
magnetoconductance ($\textrm{dMC}=dG/dH$) is large and positive
(\unit[3.0]{$\mu$S/T} at \unit[4]{K}), whereas there is essentially no
response when $\mathbf{H}$ lies in the plane (in-plane MC was measured
with $\mathbf{H}$ transverse to the current direction). This
anisotropy is found in all of our samples, and indicates that the
motion of charge carriers is confined to the graphene planes. The
observed positive dMC is in contrast to bulk graphite \cite{Soule64},
which shows negative dMC for well-ordered single-crystals, and also a
large anisotropy.  However, carbon foils fabricated from exfoliated
graphite \cite{Schaijk98} and partially graphitic carbons
\cite{Bayot90} have positive MCs initially, which become negative at
large fields.  This behavior is a consequence of disorder-induced
localization in the sample.

Figure \ref{fig:conductance}a shows systematic changes in the
perpendicular MC for samples of successively larger zero-field
conductance (i.e. decreasing disorder).  For sample C the initial
\unit[4]{K} slope is larger, at \unit[30]{$\mu$S/T}, and the MC
attains an approximately temperature-independent maximum of
\unit[185]{$\mu$S} at $H \approx \unit[3.5]{T}$. Sample A shows even
more structure.  Following a large initial dMC (\unit[500]{$\mu$S/T}
at \unit[4]{K}), the MC maximizes near \unit[1.5]{T}, then decreases,
followed by a series of (Shubnikov-de Haas) oscillations. A
temperature-independent ``fixed point'' at $G=\unit[680]{\mu S}$ and
$H=\unit[7.4]{T}$ is also observed.

As a function of temperature, the conductance increases proportional
to $\ln T$ at low $T$, as shown in Fig.~\ref{fig:conductance}b for
samples A--C. This is quite characteristic of a 2D electron gas in the
weak localization regime \cite{Abrahams79,Bergmann84}. The least
conductive samples (D and E; $G < e^{2}/h=\unit[38.8]{\mu S}$) deviate
slightly from the $\ln T$ dependence, which is indicative of a
transition to strong localization [see for instance
Ref.~\onlinecite{Minkov02}].

In a 2D system, carriers will localize at low temperature due to
constructive quantum interference of time-reversed paths for carriers
scattered elastically from static disorder \cite{Abrahams79}.  The
interference is reduced (conductance increased) by breaking
time-reversal symmetry through the application of a magnetic field, or
by increasing temperature. These coherent effects are manifest when
the elastic mean free path $\ell_e$ is smaller than the inelastic mean
free path $\ell_i(T)$. For $k_F\ell_e \gg 1$ ($k_F$ is the Fermi wave
vector), the system is in the weak localization regime. Strong
localization occurs for smaller $k_F\ell_e$ \cite{Minkov02,StrongLocal}.

Shown in the center-panel inset of Fig.~\ref{fig:conductance}a is the
perpendicular MC of sample B for five different temperatures
(circles), and fits to the data according to 2D weak-localization
theory (lines) \cite{Bergmann84}.  The entire family of MC curves is
fit by a single temperature-dependent parameter,$\ell_i(T)$.  Note
that for transport in two dimensions, the mean free paths are obtained
from the modeling without knowledge of either the Fermi velocity or
the carrier effective mass.  For sample B, we find $\ell_e =
\unit[15]{nm}$ and $\ell_i = \unit[100]{nm}$ at $T = \unit[4]{K}$.
Weak localization effects are observed over a much smaller range of
magnetic field for sample A, but a similar estimate gives $\ell_e =
\unit[\textrm{20--30}]{nm}$, and a much larger inelastic mean free
path $\ell_i(4K)\sim\unit[300]{nm}$. From the carrier density
$n=\unit[10^{12}]{cm^{-2}}$ per graphene sheet (see below), we find
for sample A $k_F\ell_{e} \sim 5$, in agreement with the weak
localization regime.

In two cases we have observed a reversal of the dMC. At the maximum MC
for sample C (Fig.~\ref{fig:conductance}a), the conductance per graphene sheet
(Table~\ref{tab:samples}) is $1.5 e^{2}/h$, i.e. comparable to the
conductance quantum. This behavior, and the large change in MC of this
sample (550\%) are reminiscent of disordered 2DEGs, which have been
explained in terms of a transition from an Anderson insulator to a
quantum Hall liquid \cite{Jiang93}.  The second case is that of sample
A, which underwent an improved substrate preparation.  For this
sample, we also observe an initial maximum in the MC, but at much
lower field (\unit[1.5]{T}).  Apparently weak-localization dominates
the MC behavior of sample A in the low-field region, but the longer
scattering paths are dephased by a smaller magnetic field.  The
subsequent appearance of Shubnikov-de Haas oscillations indicates
quantization of the electron energy spectrum, and wave function
coherence on a scale comparable to the cyclotron radius ($\sim
\unit[30]{nm}$ at \unit[5]{T}), which is consistent with the elastic
scattering lengths estimated above. The significance of the
quasi-temperature-independent fixed point at \unit[680]{$\mu$S} and a
field of \unit[7.4]{T} is not yet fully understood. Note that $1/H$ at
the crossing point is equal to the mobility obtained from Hall
measurements (see below). This correspondence would be an expected
consequence of electron-electron interactions under weak localization
\cite{Minkov02,Altshuler85}, but the observed quantum oscillations
show that at high fields the system is beyond this regime.

The Hall resistance $R_{xy}$ was measured for samples D and A in the
Hall bar configuration (photo inset, Fig.~\ref{fig:conductance}a) at a
bias current of \unit[100]{nA}. For sample D at \unit[4]{K}, $R_{xy}$
versus $H$ is linear from 0 to \unit[8]{T}, with the slope
corresponding to a density $n = \unit[10^{13}]{cm^{-2 }}$ $n$-type
charge carriers, and a mobility of \unit[15]{cm$^{2}$/Vs}.  The Hall
voltage is also linear up to \unit[4.5]{T} in sample A, from which we
determine $n = \unit[3.6\times 10^{12}]{cm^{-2 }}$ ($n$-type), and an
enhanced mobility of \unit[1100]{cm$^{2}$/Vs}.  The observation of a
linear Hall effect is particularly remarkable, since single-crystal
graphite samples display a substantial quadratic component at small
fields \cite{Schaijk98,Berlincourt55,Du04,Tokumoto04}, due to three
sub-bands (one electron, two hole).  Apparently, our samples cannot be
thus described.  The carrier densities found here are comparable to
those of other 2DEG systems, although the density per graphene sheet
($\sim \unit[10^{12}]{cm^{-2}}$) is higher than that found in
high-quality graphites \cite{Soule64,Du04,Tokumoto04}. It remains to
be determined what effect substrate doping has on the carrier density
and mobility in the graphite film.

For the case of sample A, a carrier density can be obtained from the
period (in $1/H$) of the Shubnikov-de Haas oscillations.
Assuming a circular 2D Fermi surface, we estimate $n =
\unit[10^{12}]{cm^{-2}}$ ($k_F = \unit[2.5 \times 10^{6}]{cm^{-1}}$).  The
carrier density determined by the Hall effect is very nearly a factor
3 larger, which is the number of graphene layers measured via
AES.  Clearly, the simplest explanation would be that each graphene
sheet supports a 2D electron gas which remains confined within the
sheet. This would be consistent with the large anisotropy in
conductivity for bulk graphite \cite{Soule64}. The temperature dependences of
samples A--C also support this interpretation: $dG/d(\ln T)$ falls between $2.5
e^{2}/\pi h$ and $3.5 e^{2}/\pi h$, about a factor 3
larger than the predicted weak-localization contribution to the
conductance, $3.5 (e^{2}/\pi h)\ln T$ \cite{Bergmann84}.

Above \unit[4.5]{T} in sample A, we observe nonlinearities in $R_{xy}$
vs.\ $H$, as shown in Fig.~\ref{fig:conductance}a (lower-panel inset)
for $T=\unit[4]{K}$.  These coincide with the Shubnikov-de Haas
oscillations in the magnetoconductance, showing that they have the
same origin: Either broadened quantum Hall plateaus, or bulk
magneto-quantum oscillations in a metallic system. If the Hall
conductance at the location of the local maximum
($\approx\unit[5.5]{T}$) is normalized with respect to the number of
sheets, one obtains a conductance $\approx 4 e^{2}/h$ which suggests a
quantum Hall effect (see also predictions in Ref.~\onlinecite{Zheng02}).
Experiments at lower temperatures and higher fields will be necessary
to verify this conjecture.

At \unit[0.3]{K}, we also observe a pronounced zero-bias anomaly in
the highly resistive sample D. The conductance is found to increase by
about a factor 10 as the bias voltage is increased from 0 to
\unit[25]{mV} \cite{Thermal}. For weak electron-electron interactions it can be
understood in terms of enhanced scattering of carriers near the Fermi
energy: the wavelengths of these carriers are commensurate with the
Friedel oscillations surrounding impurities, thus they scatter
strongly. The coherence is lost at higher bias (higher kinetic
energies). Zero-bias anomalies have also been observed in carbon
nanotubes \cite{Bockrath99,Tarkiainen01,Yi03}.

It should be noted that the samples are remarkably stable over time.
For instance, measurements in Figs. \ref{fig:conductance}a and
\ref{fig:conductance}b were made 4 months earlier than those in the
lower-panel inset of Fig.~\ref{fig:conductance}a, with no particular
storage precautions.  The features observed are essentially the same,
except for a slight decrease in conductance and carrier density. The
results presented above for sample A also show that the multi-sheet
epitaxial graphene film survives conventional lithographic processing
extraordinarily well.

\begin{figure}
  \includegraphics[width=0.6\columnwidth]{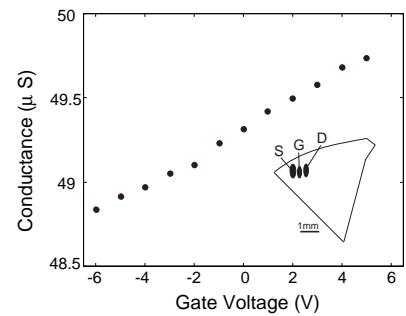}
  \caption{\label{fig:gating} Conductance as a function of gate
    voltage for sample F at \unit[4]{K}.  Inset, sketch of the sample
    showing the contacts and top-gate geometry (S, D, G = source,
    drain, gate). The top-gate is only partially effective due to the
    open geometry. Nevertheless, a 2\% change in conductance is
    observed.}
\end{figure}

Finally, as a preliminary demonstration of the device potential of
this new 2DEG system, a large-area gated graphite-channel field-effect
transistor (FET) structure was assembled.  A schematic of the
``device'' is shown in Fig.~\ref{fig:gating}, as well as the measured
source-drain resistance as a function of gate voltage. The top-gate
structure consisted of a conductive coating on a 100-nm-thick
insulating aluminum oxide layer. The gate covered only a portion of
the graphite film between the source and drain electrodes, leaving
large ungated leakage paths (see inset, Fig.~\ref{fig:gating}).
Consequently, the resistance modulation is rather small (2\%), but
these results show clearly that multi-sheet epitaxial graphene films
can be gated, in distinct contrast to thicker samples
\cite{Kempa03--1}.  Thus we anticipate that FET-type devices will be
possible, particularly when the channel electronic structure is
controlled by patterning the graphite into a narrow strip
\cite{Wakabayashi01--0,Dujardin01,Cancado04}.
%


The experimental results presented here demonstrate the rich
scientific promise of ultrathin epitaxial graphite (``graphene'')
films.  Several points should be appreciated: First, the production
method allows graphitic films to be grown epitaxially, as evidenced by
LEED and STM measurements. From Auger spectroscopy we further conclude
that the layers involve only a few graphene sheets.  Remarkably, the
films are electrically continuous over several mm.  Magnetoconductance
measurements clearly reveal 2D electron gas properties, including
large anisotropy, high mobility, and 2D localization, in samples
patterned by conventional lithography.  Quantum oscillations observed
in both the magnetoconductance and the Hall resistance indicate a
potential new quantum Hall system.  Finally, control of the 2D
electron gas carrier density via electrostatic gating was also
demonstrated.

Considered with prior research in graphitic systems, these results
provide ample evidence that the graphite/SiC system could provide a
platform for a new breed of seamlessly-integrated ballistic-carrier
devices based on nano-patterned epitaxial graphene.  Such an
architecture could have many advantages for nanoelectronics, including
potentially coherent devices, energy efficiency, and facile
integration with molecular devices.\\

\begin{acknowledgments}
This work was funded by Intel Research and by the Department of Energy
(DE-FG02-02ER45956). Support from Georgia Tech, CNRS-France, and the
NSF (ECS-0404084), is also gratefully acknowledged. We thank Drs.
Thierry Klein, Jacques Marcus and Fr\'{e}d\'{e}ric Gay, CNRS-LEPES,
Grenoble-France, for their generosity in allowing us to use their
cryostats, and Dr. P. G.  Neudeck, NASA Glenn Research Center, for
supplying the sample used in Fig.~\ref{fig:surface}. We are especially
grateful to Dr. Thierry Grenet for providing invaluable assistance in
obtaining the data in Fig.~\ref{fig:gating}.
\end{acknowledgments}


\begin{thebibliography}{10}
\bibitem{SampleNote}
Samples for transport measurements were obtained from Cree, Inc., Durham, NC
  (research grade, $n$-doped, \unit[0.05]{$\Omega$-cm}). Figures 1 and 2 show
  LEED and STM data from a sample with a 1-\unit{$\mu$m} CVD layer grown at the
  NASA Glenn Research Center. The CVD sample was initially much smoother than
  the as-received Cree sample. However, after oxidation/ graphitization cycles
  samples were found via LEED and STM to be similar, although the CVD sample
  was still of higher overall surface quality.
\bibitem{StrongLocal}
The strong localization regime is not discussed in detail here, but we note
  that for samples D and E, measurements in the range 0.3--2\unit{K} show
  $\ln(G) \sim T^{-1/3}$, a form that is consistent with theories of strong
  localization in 2D \cite{Mott87}.
\bibitem{Thermal}
Samples were carefully mounted with the connecting wires thermally connected by
  stycast epoxy to the cooled copper plate. This mounting provided more than
  sufficient thermal coupling to assure that the maximum power into the sample
  (10 nW) did not cause significant heating.
\end{thebibliography}


\end{document}